\documentclass{mn2e}
\usepackage{epsfig}

\newif\ifAMStwofonts

\newcommand{\Halpha}{H$\alpha$}

\newcommand{\Hbeta}{H$\beta$}

\newcommand{\NaD}{Na\,D}
\newcommand{\HI}{H\,{\sc i}}
\newcommand{\HeI}{He\,{\sc i}}
\newcommand{\HeII}{He\,{\sc ii}}
\newcommand{\CIII}{C\,{\sc iii}}
\newcommand{\NIII}{N\,{\sc iii}}
\newcommand{\CaII}{Ca\,{\sc ii}}
\newcommand{\target}{MM~Ser}
\title[Spectroscopy of the Optical Counterpart to Ser~X-1] 
{Spectroscopy of the Optical Counterpart to Ser~X-1}

\author[R. I. Hynes et al.]
       {R. I. Hynes$^{1,2}$\thanks{E-mail: rih@astro.as.utexas.edu;
       Hubble Fellow}
	P. A. Charles$^2$,
	L. van Zyl$^{3,4}$,
        A. Barnes$^2$,
	D. Steeghs$^{5,2}$,\newauthor
	K. O'Brien$^{6,7}$,
	J. Casares$^8$\\
$^1$Astronomy Department, The University of Texas at Austin, 1
    University Station C1400, Austin, Texas 78712-0259, USA\\
$^2$Department of Physics and Astronomy, University of Southampton, 
    Southampton, SO17 1BJ, UK\\
$^3$Department of Astrophysics, Oxford University, Keble Road, Oxford,
       OX1 3RH, UK\\
$^4$Astrophysics Group, School of Chemistry and Physics, Keele
       University, Keele, Staffordshire, ST5 5BG, UK\\
$^5$Harvard-Smithsonian Center for Astrophysics, 
    60 Garden Street, MS-67, Cambridge, MA 02138, USA\\
$^6$European Southern Observatory, Casilla 19001,
 Santiago 19, Chile\\
$^7$School of Physics and Astronomy, University of St
  Andrews, St Andrews KY16 9SS, UK\\
$^8$Instituto de Astrof\'\i{}sica de Canarias, 38200 La Laguna,
Tenerife, Spain} 
\date{Accepted ?.
      Received ?;
      in original form ?}

\pagerange{\pageref{firstpage}--\pageref{lastpage}}
\pubyear{2003}

\begin{document}
\maketitle
%
%
\begin{abstract}
We present WHT and VLT spectroscopy of \target, the optical
counterpart to Ser~X-1.  We deblend the red spectra of the two close
stars identified by Wachter (1997) and show that the brighter of the
two is responsible for the \Halpha\ and \HeI\ emission, hence
confirming that this is the true counterpart of the X-ray source.  We
also identify several \HeII\ and \NIII\ lines in the blue spectrum.
The isolated emission lines are all remarkably narrow, with FWHM
200--300\,km\,s$^{-1}$.  The Bowen blend has structure suggesting that
the individual components are also narrow.  These narrow lines could
be from the disc if the binary inclination is quite low, or they could
come from a more localised region such as the heated face of the
companion star.  Several interstellar lines are detected and indicate
that the reddening is moderate, and consistent with the neutral
hydrogen column density inferred in X-rays.
\end{abstract}
%
%
%
%
\section{Introduction}
\label{IntroSection}
The low-mass X-ray binary (LMXB) Ser~X-1 has been known as an X-ray
source from the early days of X-ray astronomy.  It was discovered in
1965 (Friedman, Byrom \& Chubb 1967).  Its optical counterpart has
proven more elusive, however.  When an accurate (1\,arcmin) position
for the X-ray source became known (Doxsey 1975), Davidsen (1975)
suggested that the optical counterpart was an ultraviolet excess
object with $B\sim18.5$.  Subsequently Thorstensen, Charles \& Bowyer
(1980) used images obtained in better seeing conditions to show that
this counterpart was actually two coincident stars (DN and DS)
separated by 2.1\,arcsec.  The southern one, designated \target, was
by far the brighter in the ultraviolet.  The detection of an optical
burst from DS, simultaneously with an X-ray burst (Hackwell et al.\
1979), confirmed that the X-ray source was associated with DS.  More
recently, however, Wachter (1997) has shown that DS is itself two
unresolved stars (DSe and DSw), separated by only 1\,arcsec.  Wachter
suggested that the brighter of the two stars, which is bluer, might be
the true optical counterpart.  Silber (1998) has analysed rapid
photometry which supports this, indicating that the brighter of the
two stars is significantly variable.  No convincing periodicity was
found in these data, however.

Spectroscopy of \target\ was obtained at several epochs, before it was
resolved into two stars.  Thorstensen et al.\ (1980) obtained blue
spectroscopy of both DS and DN. Both showed \Hbeta\ absorption, but DS
in addition had a \HeII\ 4686\,\AA\ emission line.  Cowley, Hutchings
\& Crampton (1988) confirmed the presence of narrow \HeII\ emission.
Their spectrum shows no \Hbeta\ absorption, but possible \NIII\
4640\,\AA\ emission.  Shahbaz et al.\ (1996) reported an almost
featureless continuum spanning the full optical bandpass, albeit at a
very low resolution which would be insensitive to weak emission
features.  An absorption feature around 5900\,\AA\ was attributed to a
G star secondary.

We report on further spectroscopy of \target.  Our primary goal was to
separate the spectra of DSe and DSw and hence confirm which is the
true optical counterpart to Ser~X-1.  Since the most recent published
spectroscopy dates from 1988 we also were able to obtain a higher
quality spectrum using modern instrumentation and hence study the
properties of \target\ in more detail than previously possible.

%
\section{Observations}

\subsection{WHT spectra}
\label{WHTSection}
Spectra of \target\ were obtained from the Observatorio del Roque de
los Muchachos on La Palma on the nights of 2001 July 8 and 10 using
the 4.2\,m William Herschel Telescope (WHT) with the ISIS dual-arm
spectrograph.  The seeing was typically 1.0--1.5\,arcsec on July 8 and
$\sim0.8$\,arcsec on July 10.

The slit width was set to 0.83\,arcsec.  The slit was aligned to pass
through the two blended stars (DSe and DSw) identified by Wachter
(1997).  The blue arm was used with the EEV12 CCD and the R600B
grating to yield a resolution of 1.9\,\AA.  Note that the blue
spectrum is heavily vignetted at either end; the useful coverage spans
3720--5260\,\AA.  On July 8 the red arm was used with the TEK4 CCD and
the R600R grating for a resolution of 1.5\,\AA.  On July 10 the R316R
grating was instead used giving a resolution of 3.1\,\AA.  A log of
the spectra obtained of \target\ is given in Table~\ref{ObsTable}.

\begin{table*}
\caption{Log of spectroscopic observations of \target.}
\label{ObsTable}
\begin{center}
\begin{tabular}{llllcccc}
\hline
Date & Telescope & Start     & End       & Number of & Exp.\    & Wavelength &
Resolution \\
     & & time (UT) & time (UT) & exposures & time (s) & range (\AA) &
(\AA) \\
\noalign{\smallskip}
2001 July 8  & WHT/ISIS & 22:09 & 23:54 & 4 & 1500 & 3600--5390 & 1.9 \\
2001 July 8  & WHT/ISIS & 22:09 & 23:54 & 4 & 1500 & 5870--6660 & 1.5 \\
2001 July 10 & WHT/ISIS & 21:47 & 22:47 & 2 & 1800 & 3600--5390 & 1.9 \\ 
2001 July 10 & WHT/ISIS & 21:47 & 22:48 & 2 & 1800 & 5810--7300 & 3.1 \\
\noalign{\smallskip}
2002 May 18 & VLT/FORS2 & 08:29 & 09:25 & 3 & 1100 & 4510--5810 & 1.8 \\
\noalign{\smallskip}
\hline
\end{tabular}
\end{center}
\end{table*}

Basic image processing (bias correction and flat fielding) was done
using standard {\sc iraf}\footnote{IRAF is distributed by the National
Optical Astronomy Observatories, which are operated by the Association
of Universities for Research in Astronomy, Inc., under cooperative
agreement with the National Science Foundation.} techniques.
Unfortunately no suitable red arm flat field was obtained on July 10,
so it was necessary to use the July 8 one which was obtained with a
different grating.

Initially a combined spectrum of both stars was extracted using the
{\sc iraf} implementation of optimal extraction (Horne 1986; Marsh
1989).  As there are actually two blended stars at the site of
\target, this will not truly be an optimal extraction in this case.
The spectra extracted in this way were still far superior to a
non-optimal extraction, however.

Wavelength calibration was interpolated relative to exposures of a
CuNe+CuAr lamp and checked against night sky lines.  The rms scatter
among the latter was 0.25\,\AA\ in the blue and 0.14\,\AA\ in the red,
with no systematic offset in either case.

Flux calibration was applied relative to the spectrophotometric
standard BD+33$^{\circ}$2642 (Oke 1990); this was also used to correct
for Telluric absorption features.  As an 0.8\,arcsec slit was used
significant uncertainty will be introduced in the flux calibration due
to slit losses, especially as the slit was not aligned at the
parallactic angle due to the need to follow the line of centres of the
blended star.  Consequently, the flux calibration of the spectra is
only approximate.

\begin{figure}
\begin{center}
\epsfig{angle=90,width=3.4in,file=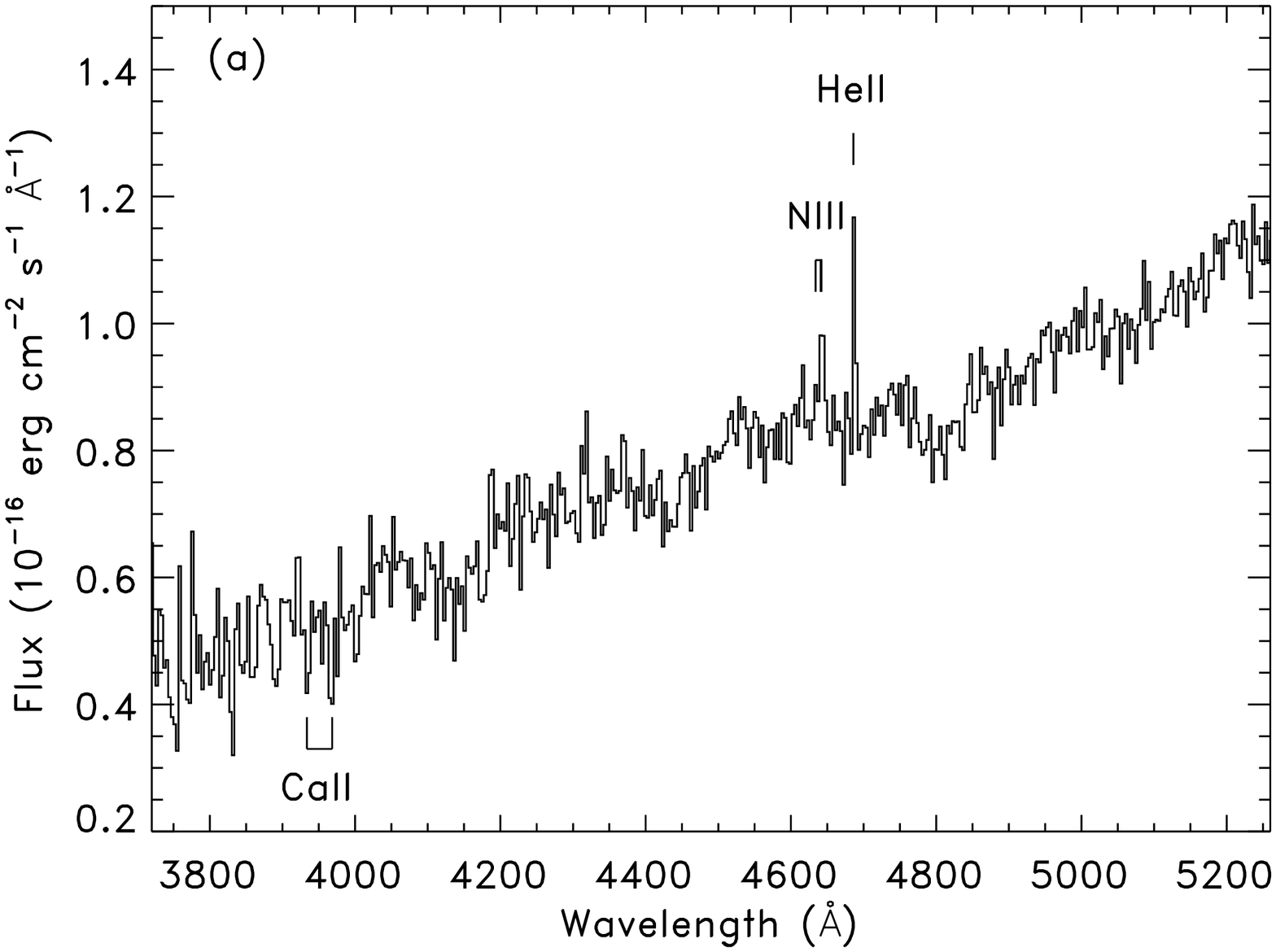}
\epsfig{angle=90,width=3.4in,file=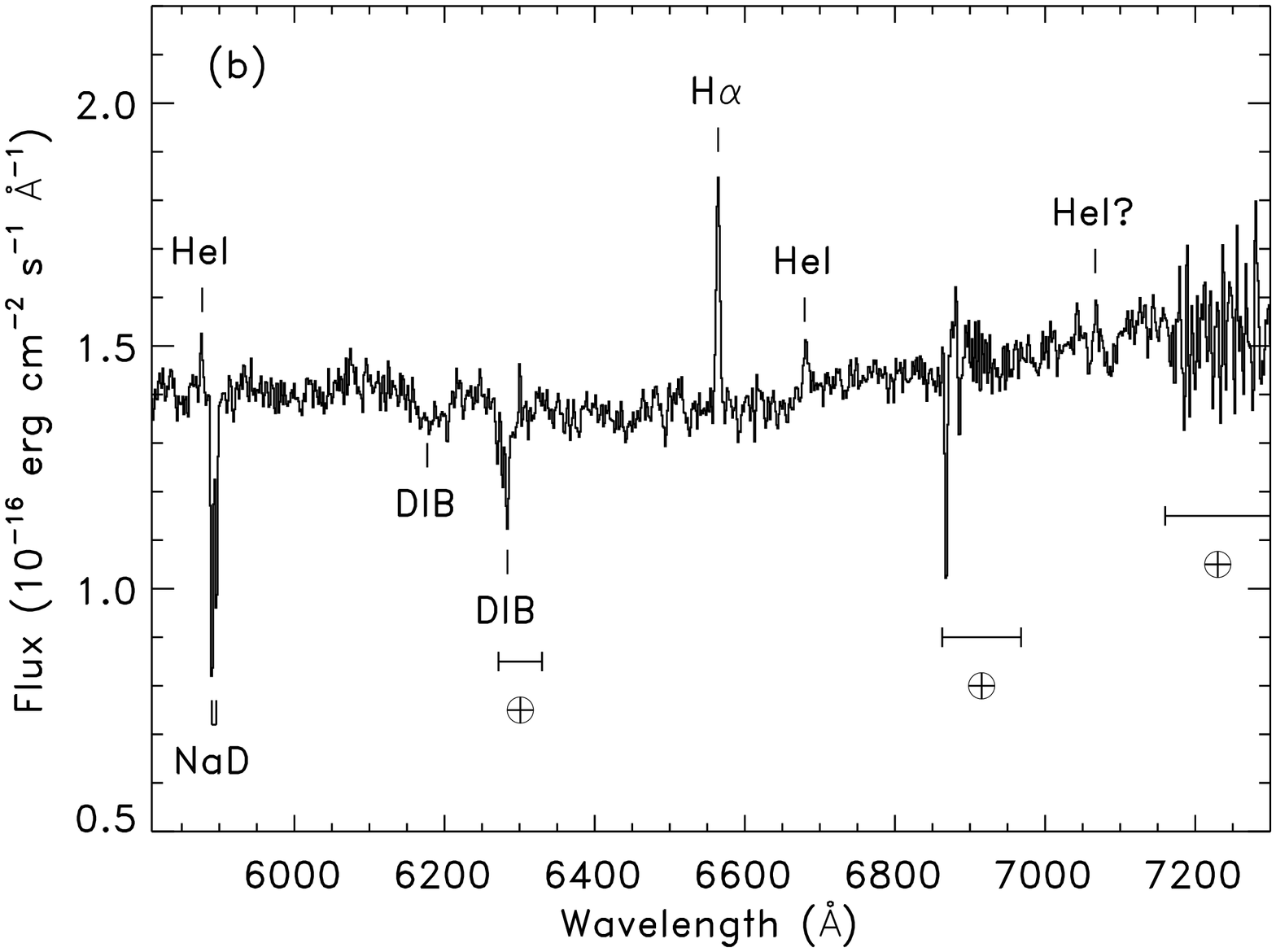}
\caption{{\bf a)} Combined blue spectrum of the two stars, binned to
  3.5\,\AA\ pixels.  This will be dominated by \target\ with the other
  star contributing less than 20\,percent of the light.  
  {\bf b)} Combined red spectrum.  Identified emission and interstellar
  absorption features are marked.  Telluric correction has been
  applied, but regions which may be contaminated by residuals from
  this process are marked $\oplus$.  The Telluric feature around
  6300\,\AA\ is much weaker than the longer wavelength features, so
  this region is probably not significantly contaminated.}
\label{AvSpecFig}
\end{center}
\end{figure}

The average extracted spectra (a composite of the two blended stars)
is shown in Fig.~\ref{AvSpecFig}.  The presence of {\em strong}
emission lines in the combined spectrum suggests that the brighter,
bluer star is the optical counterpart to \target\ as suggested by
Wachter (1997).  To confirm this we also used the spectral deblending
method of Hynes (2002) to extract spectra of the two stars separately.

\subsection{VLT Spectra}
\label{VLTSection}
Three further spectra were obtained with the FORS2 spectrograph at the
Very Large Telescope (VLT) on 2002 May 18.  The 1400V grism was used
with an MIT/Lincoln Labs mosaic CCD and an 0.7\,arcsec slit.  The
seeing was around 1.2\,arcsec. Further details are given in
Table~\ref{ObsTable}.

\begin{figure}
\begin{center}
\epsfig{angle=90,width=3.4in,file=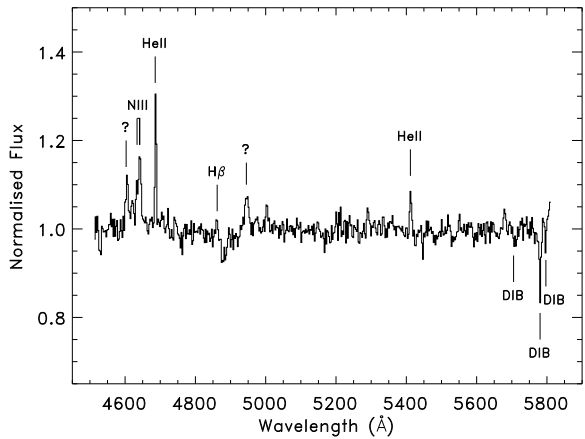}
\caption{Average, normalised VLT spectrum of \target.  Identified
  features are marked. Unidentified features appear to be present at
  4605 and 4945\,\AA.  These are repeatable, and not obviously
  artifacts.}
\label{VLTSpecFig}
\end{center}
\end{figure}

Images were debiased and flatfielded using Starlink {\sc ccdpack}
routines.  As no suitable PSF template was available, and the
contribution of the fainter star appeared very small, we only
performed a single object `optimal' extraction with {\sc iraf} as
above.  The images were marred by a number of bright spots, typically
up to 20--30\,pixels across.  These originate from enhanced dark
current in regions that had been heavily saturated during earlier
observations.  We examined the saturated images to identify which
regions of the spectrum might be affected.  Fortunately, there were
only a few mildly saturated regions coincident with the location of
our spectra, and none of these correspond to apparent features in the
extracted one-dimensional spectra.  We therefore believe that this
problem does not compromise our dataset.  Wavelength calibration was
applied with respect to an arc image taken during the day.  The
wavelength was checked with respect to sky lines (mainly [O\,{\sc i}]
5577\,\AA).  There was an offset of $+0.21$\,\AA\ (13\,km\,s$^{-1}$ at
4686\,\AA) which was corrected.  This is within the expected flexure
for FORS2 (Szeifert 2002).  Given the large slit losses due to poor
seeing, no flux calibration of this spectrum was attempted.  The final
average spectrum is shown in Fig.~\ref{VLTSpecFig}.

%
\section{Identification of the counterpart}
\label{DeblendSection}
\begin{figure}
\begin{center}
\epsfig{angle=90,width=3.4in,file=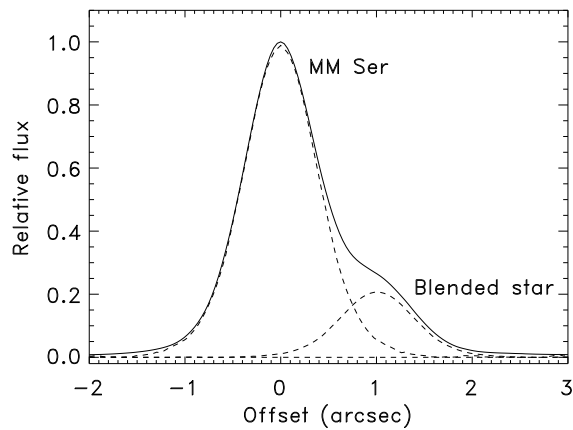}
\caption{Average WHT red spatial profile showing \target\ and the
  blended star.  The solid line shows the data, oversampled and
  averaged over wavelength after removing the curvature of the
  spectrum.  The dashed lines show the corresponding averages of the
  two modelled profiles.}
\label{SpatialProfileFig}
\end{center}
\end{figure}

\begin{figure}
\begin{center}
\epsfig{angle=90,width=3.4in,file=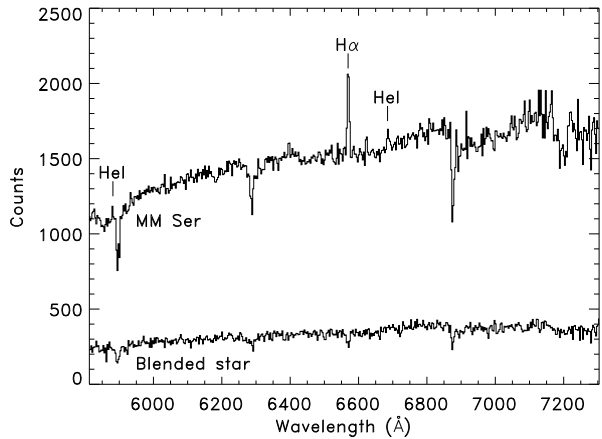}
\caption{Average of the two deblended red spectra of \target\ and the
nearby star from 2001 July 10.  It is clear that the emission lines,
particularly \Halpha, are associated with the brighter of the two
stars.}
\label{DeblendFig}
\end{center}
\end{figure}

To separate the two stars we applied the deblending algorithm of Hynes
(2002) to the July 10 red WHT spectra.  These had the best spatial
resolution and were well exposed, and were the only spectra suitable
for this work.  The stars were not fully resolved
(Fig.~\ref{SpatialProfileFig}), but were sufficiently separated that
they could be identified and deblended.  We defined the spatial
profile with a base Voigt function, with width allowed to vary with
wavelength, together with a wavelength independent numerical
correction.  The deblended spectra are shown in Fig.~\ref{DeblendFig}.

It is clear that the emission features detected do originate from the
bright star.  \Halpha\ is particularly striking, but \HeI\ 6678\,\AA\
is also recognisable.  These features are clear in both individual
spectra as well as in the average.  This provides further confirmation
of earlier suggestions (Wachter 1997; Silber 1998) that the brighter
of the two stars, DSe in the terminology of Wachter (1997), is the
LMXB.

Comparing the two components, at the epoch observed the fainter star
contributes only about 15--20\,percent of the total light in the
6000-7000\,\AA\ band.  For comparison, the magnitudes given by Wachter
(1997) correspond to contributions of 19\,percent at $V$ and
25\,percent at $R$.

%
\section{The optical spectrum of \target}
\label{SpecSection}
\subsection{Emission lines}
The only emission lines clearly detected in the WHT blue spectrum are
\HeII\ 4686\,\AA\ and the \NIII\ Bowen blend around 4640\,\AA; there
are no strong Balmer emission lines in the blue.  The Bowen blend
contains at most weak \CIII\ emission.  Moving to the red, the WHT
spectrum is dominated by \Halpha, with \HeI\ 5875 and 6678\,\AA\ also
present.  The VLT spectrum partially overlaps the WHT blue-side, but
also fills in coverage around 5500\,\AA.  It shows the same blue
lines, of \NIII\ and \HeII\ 4686\,\AA, with a similar relative
strength.  The \HeII\ 5411\,\AA\ line is also clearly present.  Weak
\Hbeta\ emission does appear to be present, on the blue-side of a
broad absorption trough.  This is a ubiquitous feature in both black
hole and neutron star LMXB spectra, e.g.\ Buxton \& Vennes (2003); see
also Soria et al.\ (2000) and references therein for other black hole
examples and Hynes et al.\ (2001) and Casares et al.\ (2003) for other
neutron star cases.  If it is indeed associated with \Hbeta, the large
redshift ($\sim1000$\,km\,s$^{-1}$) remains unexplained.  It is
possible that the apparent redshift is only due to filling in of the
blue absorption by emission, however.  Such absorption troughs are
usually attributed to the inner optically thick accretion disc (e.g.\
Soria et al.\ 2000), but a large redshift would not then be expected.
Apparent features are seen at 4605 and 4945\,\AA.  They do not appear
to be artifacts, but there is no obvious identification and no strong
features are seen in other LMXBs at these wavelengths.  Given the
inferior quality of the WHT spectra, they would probably be
undetectable there.  Properties of lines measured in all of the
spectra are summarised in Table~\ref{LineTable}.

\begin{table*}
\caption{Measured line properties.}
\label{LineTable}
\begin{tabular}{lllll}
\noalign{\smallskip}
\hline
\noalign{\smallskip}
Date & Line & LSR Velocity & EW & FWHM \\
     &      & (km\,s$^{-1}$) & (\AA) & (km\,s$^{-1}$)\\
\noalign{\smallskip}
2001 July 8--10 &\NIII\ 4634,41,42 & -- & $1.9\pm0.4$ & -- \\
(WHT/ISIS)      &\HeII\ 4686\,\AA & $+127\pm11$ & $2.1\pm0.2$ & $200\pm30$ \\
                &\HI\ 6562\,\AA\ (\Halpha) &  $+90\pm6$  & $2.1\pm0.1$ & $210\pm20$ \\
                &\HeI\ 6678\,\AA & $+165\pm40$ & $0.5\pm0.1$ & $280\pm80$ \\
\noalign{\smallskip}
2002 May 18 &\NIII\ 4634,41,42 & -- & $1.6\pm0.4$ & -- \\
(VLT/FORS2)  &\HeII\ 4686\,\AA & $+112\pm6$  & $1.4\pm0.1$ & $240\pm20$ \\
            &\HeII\ 5411\,\AA &  $+42\pm16$ & $0.6\pm0.1$ & $320\pm40$ \\
\noalign{\smallskip}
\hline
\end{tabular}
\end{table*}

\begin{figure}
\begin{center}
\epsfig{width=3.4in,file=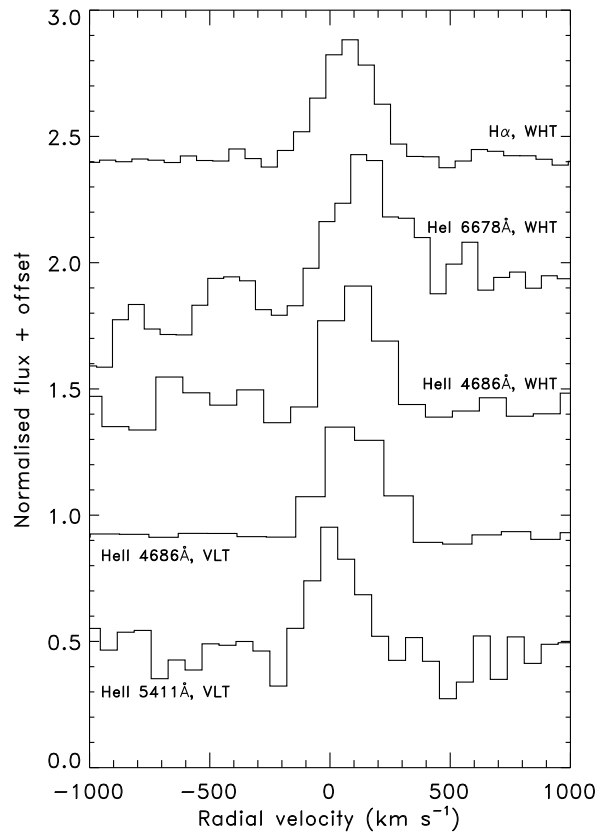}
\caption{Line profiles of the main, identified, non-blended lines.
  All are rather narrow and appear single-peaked, at least to the
  limits of the resolution.  Spectra have been binned to approximately
  one pixel per resolution element in each case.  They have been
  continuum subtracted and normalised to the same total line flux, so
  that the shapes of the profiles can be compared.}
\label{ProfileFig}
\end{center}
\end{figure}

All of the lines are remarkably narrow, as can clearly be seen in
Fig.~\ref{ProfileFig}, where the line profiles in velocity space are
collated.  The FWHM listed in Table~\ref{LineTable} are based on
single Gaussian fits to the unbinned spectra.  The contribution from
the instrumental resolution has been corrected for, but is never
dominant.  For comparison the FWHM of \HeII\ in 4U\,1822--371 (Casares
et al.\ 2003) is FWHM 750\,km\,s$^{-1}$, and the lines in
4U\,1735--444 and 4U\,1636--536 respectively have FWHMs of
$660$\,km\,s$^{-1}$ and 1220\,km\,s$^{-1}$ (Casares et al.\ in
preparation).  Such narrow emission lines are thus rather atypical for
an LMXB.  If they are disc lines, then the narrowness could indicate
either a long period, and hence a large disc with a low outer
velocity, or a low inclination, or a combination of both.  Without
fully resolving the profiles and confirming that they are disc lines,
it is of limited value to model them.  Applying some simple models to
the \HeII\ 4686\,\AA\ line, however, suggests that for typical neutron
star LMXB parameters, if the period is less than 1\,day then
$i\la20^{\circ}$.  A longer period allows a somewhat higher
inclination; for 10\,days then $i\la45^{\circ}$.  These are not very
robust estimates, however, as when system parameters are known,
apparent sub-Keplerian velocities are often inferred from disc line
profiles (Marsh 1998).  Alternatively the lines may not be dominated
by the disc; they could come from the companion star or stream-impact
point, or some other region (as in XTE~J2123--058; Hynes et al.\
2001), in which case the narrowness would be more natural.  In this
case, unless the inclination were low, we would expect the lines to
move.  There is clearly no large velocity difference between the two
epochs, and the \HeII\ 4686\,\AA\ velocities measured are consistent
within errors, especially allowing for some uncertainty in the
absolute velocities.  The WHT spectra are individually too poor to
search for shorter timescale radial velocity variations.  The three
VLT spectra are better suited.  They are equally spaced and the
5577\,\AA\ night sky line is stable to $\pm0.01$\,\AA\ between them.
We measure Gaussian fit velocities for \HeII\ 4686\,\AA\ in the
individual spectra of $105\pm13$\,km\,s$^{-1}$,
$111\pm10$\,km\,s$^{-1}$, and $133\pm9$\,km\,s$^{-1}$ respectively.
These suggest a small drift (over about an hour), but this is not
significant at the 90\,percent confidence level and more observations
would be needed to confirm any motion.

\begin{figure}
\begin{center}
\epsfig{angle=90,width=3.4in,file=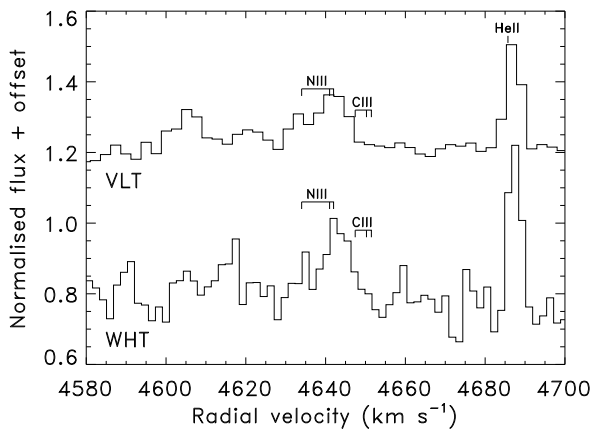}
\caption{The spectral region containing the \NIII\ 4634/41/42\,\AA\
  blend and \HeII\ 4686\,\AA.  All annotations indicate the laboratory
  wavelength of expected lines.  \HeII\ is clearly somewhat
  redshifted, and \NIII\ may be as well.}
\label{BowenFig}
\end{center}
\end{figure}

The region around the \NIII\ lines and \HeII\ 4686\,\AA\ is expanded
and compared in Fig.~\ref{BowenFig}.  The \NIII\ blend appears to show
repeating structure with a strong peak corresponding to the
4641/42\,\AA\ lines and a weaker peak for 4634\,\AA.  There is no
evidence for \CIII\ 4647/50/51 emission.  That these components are
resolved suggests that the individual components are narrow, like the
isolated lines.  The ratio of the Bowen and \HeII\ equivalent widths
is $\sim0.8$, typical of Galactic plane LMXBs which have values of
0.5--1.0 (Motch \& Pakull 1989).

\subsection{Interstellar features}
Our spectra show a number of interstellar features; \NaD\ lines,
diffuse interstellar bands (DIBs) and perhaps Ca\,H and K lines.  We
can use these to constrain the amount of reddening toward \target,
although since these measures depend on measurements of equivalent
widths, it is only valid to use measures from the deblended red
spectrum.  Shahbaz et al.\ (1996) suggested that the strong \NaD\
feature in the spectrum of \target\ indicated a G-type companion star.
Since the spectrum shows strong diffuse interstellar bands (DIBs), we
instead feel that it is more likely that the \NaD\ absorption is also
of interstellar origin.  \CaII\ H and K lines (3933,3968\,\AA) are
also marginally detected in the blue spectrum, which should be
dominated by the accretion light; hence these favour an interstellar
origin.  The interstellar absorption is therefore likely large.

The \NaD\ doublet can be used as a reddening indicator, but is only
reliable for small reddenings ($E(B-V) \la 0.5$), as it is prone to be
saturated at larger values (Munari \& Zwitter 1997).  We measure an EW
for the \NaD\ 5890\,\AA\ line of $1.6\pm0.1$\,\AA\ from the deblended
spectrum of DSe.  This is larger than any values given by Munari \&
Zwitter (1997), which saturate at a maximum value of about 1.2\,\AA.
There is a spread in the saturation values, however, as it depends on
the relative velocities of the clouds contributing to the absorption.
A larger maximum \NaD\ EW is seen when multiple resolved components
are present as this partly circumvents saturation.  Since all of the
stars considered by Munari \& Zwitter (1997) are at $<3$\,kpc and
\target\ has an estimated distance of 8.4\,kpc (Christian \& Swank
1997), it is likely that a larger EW is possible in the latter, since
\NaD\ absorption will be spread over a wider range of velocities and
hence be less affected by saturation.  We can still use the \NaD\
strength as a lower limit for the reddening (if it is of interstellar
origin), since the asymptotic linear relation between EW and $E(B-V)$
at low reddenings defines the limiting case of unsaturated absorption.
The relation given by Munari \& Zwitter (1997) will then define a
lower limit of of $E(B-V) \ga 0.4$.  We note that the ratio of the
lines is also an indicator of reddening (Munari \& Zwitter 1997),
since the \NaD\ 5896\,\AA\ line is intrinsically weaker and so
saturates more slowly.  At low optical depths the ratio is 2.0, but
this decreases to 1.1 as the reddening increases.  We measure a ratio
of $1.3\pm0.1$ suggesting that at least some components are saturated
and hence that the reddening is high, significantly larger than the
lower limit.

A relatively high reddening is supported by the strong DIBs; from the
DIB at 6203\,\AA\ we measure an EW of $270\pm130$\,m\AA, implying
$E(B-V)=0.8\pm0.4$ (Herbig 1975).  The DIB at 6283\,\AA\ may also be
strong, but will not be reliable due to Telluric contamination.  This
value is consistent with the column density of $N_{\rm H} =
0.5\times10^{21}$\,cm$^{-2}$ estimated by Christian \& Swank (1997),
since the latter would imply $E(B-V)=0.6-1.1$ (e.g.\ Bohlin et al.\
1978; Predehl \& Schmidtt 1995).  We conclude that \target\ is
moderately reddened, with all indicators consistent with a reddening
$E(B-V)\sim0.6-1.1$.
%
%
\section{Conclusions}
\label{ConclusionSection}
We have performed spectroscopy of \target, the optical counterpart to
Ser~X-1.  We resolve the two close components identified by Wachter
(1997) and confirm earlier suggestions that the brighter of the two is
an emission line source and hence the true counterpart.  The spectra
reveal emission lines of \HI, \HeI, \HeII, and \NIII, as well as
interstellar features.  The emission lines are unusually narrow.
Since all lines show similar widths, the most likely explanation seems
that they are disc lines, but that the binary inclination is very low.
Alternatively they could also originate from a more localised region
such as the companion star or stream-impact point, but we would then
require that the disc not dominate any of the optical lines.  Further
time-resolved observations would be needed to search for motion of the
lines and discriminate between these possibilities.  Higher resolution
spectroscopy would also be beneficial to resolve the narrow line
profiles better and study the substructure of the \NIII\ blend.  Based
on the strong interstellar lines, we have estimated a reddening
consistent with estimates based on the X-ray derived neutral hydrogen
column density of $E(B-V)=0.6-1.1$.
%
%
\section*{Acknowledgements}
We are grateful to Carole Haswell for providing details of Silber
(1998), and to Hilmar Duerbeck for permission to examine his
proprietary VLT images.  RIH and PAC acknowledge support from grant
F/00-180/A from the Leverhulme Trust.  RIH is currently funded from
NASA through Hubble Fellowship grant \#HF-01150.01-A awarded by the
Space Telescope Science Institute, which is operated by the
Association of Universities for Research in Astronomy, Inc., for NASA,
under contract NAS 5-26555.  LvZ acknowledges the support of
scholarships from the Vatican Observatory, the National Research
Foundation (South Africa), the University of Cape Town, and the
Overseas Research Studentship scheme (UK).  DS acknowledges the
support of a PPARC Postdoctoral Fellowship and a Smithsonian
Astrophysical Observatory Clay Fellowship.  The William Herschel
Telescope is operated on the island of La Palma by the Isaac Newton
Group in the Spanish Observatorio del Roque de los Muchachos of the
Instituto de Astrof\'\i{}sica de Canarias.  The Very Large Telescope
is operated by the European Southern Observatory at Cerro Paranal in
Chile.  This research has made use of the SIMBAD database, operated at
CDS, Strasbourg, France and the NASA Astrophysics Data System Abstract
Service.
%
%

%
\end{document}